\documentstyle[12pt]{article}
\begin{document}
\topmargin -1cm
\centerline{\Large {\bf Distributions of Time Headways}}
\centerline{\Large {\bf in Particle-hopping Models of Vehicular Traffic}} 
\vspace{1cm}
\centerline{\bf Kingshuk Ghosh, Arnab Majumdar and Debashish Chowdhury$^1$}
\centerline{Physics Department, Indian Institute of Technology} 
\centerline{Kanpur 208016, India}
\vspace{.5cm}
\renewcommand{\baselinestretch}{2}\huge\normalsize
\begin{abstract}
We report the first analytical calculation of the distribution of the 
time headways in some special cases of a particle-hopping model of 
vehicular traffic on idealized single-lane highways and compare with the 
corresponding results of our computer simulation. We also present numerical 
results for the time-headway distribution in more general situations in 
this model. We compare our results with the empirical data available in 
the literature. 

\end{abstract}

\noindent PACS. 05.40.+j - Fluctuation phenomena, random processes,
and Brownian motion.

\noindent PACS. 05.60.+w - Transport processes: theory.

\noindent PACS. 89.40.+k - Transportation.

-----------------------------------------------------------------------------------------

$^1$To whom all correspondence should be addressed; E-mail: debch@iitk.ernet.in

\vfill\eject 

Significant progress has been made in recent years in understanding the 
dynamical phases of vehicular traffic using $particle-hopping$ models [1]; 
these are analogous to the "microscopic" models of interacting classical 
particles studied in statistical mechanics.  
The time interval between the arrivals at (or departures from) a detector site 
of two successive vehicles is defined as the $time$ $headway$ between this 
pair of vehicles [2]. The distribution of time headways is one of the most 
important characteristics of vehicular traffic and the aim of this letter is 
to calculate this distribution, for a wide range of densities of vehicles, 
in the steady state of the Nagel-Schreckenberg (NS) model [3] of traffic on 
highways. 

In the NS model a lane is represented by a one-dimensional lattice of 
$L$ sites. Each of the lattice sites can be either empty or occupied by at 
most one "vehicle". If periodic boundary condition is imposed, the 
density $c$ of the vehicles is $N/L$ where $N (\leq L)$ is the total number 
of vehicles. 

In the NS model [3] the speed $V$ of each vehicle can take one of the 
$V_{max}+1$ allowed {\it integer} values $V=0,1,...,V_{max}$. 
At each {\it discrete time} step $t \rightarrow t+1$, the arrangement of $N$ 
vehicles is updated {\it in parallel} according to the following "rules".
Suppose, $V_n$ is the speed of the $n$-th vehicle at time $t$. 

{\it Step 1: Acceleration.} If, $ V_n < V_{max}$, the speed 
of the $n$-th vehicle is increased by one, i.e., $V_n \rightarrow V_n+1$.

{\it Step 2: Deceleration (due to other vehicles).} If $d$ is the gap
in between the $n$-th vehicle and the vehicle in front of it, and if 
$d \le V_n$, the speed of the $n$-th vehicle is reduced to $d-1$, i.e., 
$V_n \rightarrow d-1$.

{\it Step 3: Randomization.} If $V_n > 0$, the speed of the $n$-th vehicle 
is decreased randomly by unity (i.e., $V_n \rightarrow V_n-1$) with 
probability $p$ ($0 \leq p \leq 1$); $p$, the random deceleration probability, 
is identical for all the vehicles and does not change during the updating.

{\it Step 4: Vehicle movement.} Each vehicle is moved forward so that 
$X_n \rightarrow  X_n + V_n$ where $X_n$ denotes the position of the 
$n$-th vehicle at time $t$.\\  

We shall present our analytical calculations for $V_{max} = 1$ and compare 
these with the corresponding results of our computer simulation. As analytical 
calculations are too complicated to carry through for $V_{max} > 1$, we have 
computed the time-headway distributions for all $V_{max} > 1$ only through 
computer simulation. 

For the convenience of our analytical calculations, we change the order of 
the steps in the update rules in such a manner that does not influence the 
steady-state properties of the model [4]. We assume the sequence of steps 
$2-3-4-1$, instead of $1-2-3-4$; the advantage is that there is no vehicle 
with $V = 0$ immediately after the acceleration step. Consequently, if 
$V_{max}=1$, we can then use a binary site variable $\sigma$ to describe the 
state of each site; $\sigma=0$ represents an empty site and $\sigma = 1$ 
represents a site occupied by a vehicle whose speed is unity. 
   
We label the position of the detector by $j=0$, the site immediately in 
front of it by $j=1$, and so on. The detector clock resets to $t=0$ everytime 
a vehicle leaves the detector site. We begin our analytical calculations 
for $V_{max} = 1$ by {\sl assuming} that the probability of a time headway 
$t$, ${\cal P}(t)$, between a "leading" vehicle ({\bf LV}) and the "following" 
vehicle ({\bf FV}) is given by
\begin{equation}
  {\cal P}(t) = \sum_{t_1=1}^{t-1} P(t_1)Q(t-t_1) 
\end{equation}
where $P(t_1)$ is the probability that there is a time interval $t_1$ 
between the departure of the LV and the arrival of the FV at the detector 
site and $Q(t_2)$ is the probability that the FV halts at the detector 
site for $t_2$ time steps. 

       To calculate $P(t_1)$, we have to consider spatial configurations 
       at $t=0$ for which the FV can reach the detector site 
	   within $t_1$ steps.
       This means considering configurations upto the maximum seperation 
       of $t_1$ sites from the detector site so that even the farthest 
	   vehicle can reach the detector site with $t_1$ hops. 
       The configurations of interest are thus of the form
$$ (\underbrace{100\cdots0}_{n}|\underline{0}) \quad n=1,2,\cdots t_1.  $$ 
       The underlined zero implies that we have to find the conditional 
       steady-state probability for the given configuration subject to
       the condition that the underlined site (detector site) is empty.
       This probability, $\Pi(n)$, in the 2-cluster approximation is given 
	   by [4,5]
       \begin{equation}
         \Pi(n) = {\cal C}(1|\underline{0})\left\{ {\cal C}(0|\underline{0})
		  \right\}^{n-1}, 
       \end{equation}
       where, ${\cal C}$ gives the 2-cluster steady-state configurational 
       probability for the argument configuration and the underlined imply
       the conditional, as usual. The expressions for the various ${\cal C}$s
       are given by [4,5] 
       \begin{equation}
         {\cal C}(0|\underline{0}) \quad =\quad {\cal C}(\underline{0}|0) = 1 - \frac{y}{d} 
       \end{equation}
       \begin{equation}
       {\cal C}(0|\underline{1}) \quad =\quad {\cal C}(\underline{1}|0) = \frac{y}{c} 
       \end{equation}
       \begin{equation}
       {\cal C}(1|\underline{0}) \quad =\quad {\cal C}(\underline{0}|1) =  \frac{y}{d} 
       \end{equation}
       \begin{equation}
       {\cal C}(1|\underline{1}) \quad =\quad {\cal C}(\underline{1}|1) = 1 - \frac{y}{c} 
       \end{equation}
       where 
       \begin{equation}
         y = \frac{1}{2q}\left( 1 - \sqrt{1 - 4 q c d}\right), 
       \end{equation} 
	   $q = 1 - p$ and $d = 1 - c$.
       
       For all configurations with $t_1>n$, $t_1-1$ time steps elapse in 
	   crossing $n - 1$ bonds (as the last bond is crossed certainly 
	   at the last time step). Thus 
       \begin{equation}
	 P(t_1) = \sum_{n=1}^{t_1} \Pi(n) q^n p^{t_1-n} \quad ^{t_1-1}C_{n-1} 
       \end{equation}
    Using the expression (2) for $\Pi(n)$ in equation (8) we get 
\begin{equation}
	  P(t_1)  =  {\cal C}(1|\underline{0})q
	\left[{\cal C}(0|\underline{0})q + p \right]^{t_1-1} 
\end{equation}

       Next we calculate $Q(t_2)$ by considering all configurations of the form
       \[ (\underline{1}|\underbrace{11\cdots 1}_{m-1}0) \]  at $t=t_1$.
       Here, the underlined part implies that the detector site is occupied 
	   by the FV and we consider configurations with a queue 
	   of ($m-1$) vehicles ahead of it and the foremost vehicle of this queue 
	   has an empty site ahead. The steady-state probability of this 
	   configuration is given by
\begin{equation}
 \Pi^{\prime}(m) = {\cal C}(\underline{1}|\underbrace{11\cdots 1}_{m-1}0)  = \left\{ {\cal C}(\underline{1}|1) \right\}^{m-1}{\cal C}(\underline{1}|0)
\end{equation} 
in 2-cluster approximation. To find all possible configurations of this type 
that contribute to the waiting time of $t_2$ time-steps, we have to consider 
queue sizes upto $t_2$ ($m\le t_2$). Thus, using expression (10) for 
$\Pi^{\prime}(m)$ in 
$$ Q(t_2) = \sum_{m=1}^{t_2} \Pi^{\prime}(m) q^m p^{t_2-m} \quad ^{t_2 - 1}C_{m-1}$$ 
we get  
\begin{equation}
Q(t_2)  = {\cal C}(\underline{1}|0) q \left[ q {\cal C}(\underline{1}|1) + p  \right]^{t_2 -1}. 
\end{equation}
       
The time-gap distribution, ${\cal P}(t)$, can now be calculated using equations 
(1), (9) and (11) and is given by 
\begin{eqnarray}
{\cal P}(t) = q^2 {\cal C}(1|\underline{0}){\cal C}(\underline{1}|0) \left[
          \frac{\left[{\cal C}(0|\underline{0})q + p \right]^{t-1}
          -\left[ q {\cal C}(\underline{1}|1) + p \right]^{t-1}}
          {\left[{\cal C}(0|\underline{0})q + p \right]
          -\left[ q {\cal C}(\underline{1}|1) + p \right]} \right] \nonumber \\ \label{eq-app1}
          = \frac{yq}{c-d} \left[ \left( 1 - \frac{yq}{c}\right)^{t-1} - \left( 1 - \frac{yq}{d}\right)^{t-1} \right]
\end{eqnarray}
where the final result (eq.(12)) was obtained using equations~(3)-(6) and y is 
given by eq.~(7). 
	   
In the approximate calculation of $Q(t_2)$ we ignored the fact that, for 
$V_{max} = 1$, the LV is {\it certainly} present at $j=1$ at $t=0$.  
If the LV is still at site $j=1$ at $t > t_1$, when the FV is at $j=0$, it 
hinders the forward movement of the FV; therefore, we now recalculate 
${\cal P}(t)$ incorporating leading corrections. The LV itself may be 
hindered by its predecessor halting at site $j=2$ and we get a hierachy of 
terms. We {\sl assume} that the probability of a $m$-sized queue ahead of, 
and including the LV, is given by an expression identical to equation~(10). 
Using arguments similar to the one used earlier for deriving equation~(11), 
one can show that the probability of the LV hopping out of site $j=1$ at 
$t=t_3$ is given by $Q(t_3)$ (equation~(11)). The expression (9) for $P(t_1)$ 
remains unaffected but the equation~(1) gets modified by this correction. 
The expression for the total probability ({\sl i.e.} ${\cal P}(t)$) is now 
given by 
\begin{equation}
{\cal P}(t) = \sum_{t_1=1}^{t-1} \left\{ q p^{t-t_1-1} P(t_1) S(t_1) + q P(t_1) R(t_1,t-t_1) \right\} 
\end{equation}
where 
\begin{equation}
S(t_1) =  Q(1) + Q(2) + \cdots + Q(t_1) = \sum_{t_3=1}^{t_1} Q(t_3)
\end{equation}
and 
\begin{equation}
R(t_1,t_2) =  Q(t_1 +1) p^{t_2-2} + Q(t_1 +2) p^{t_2-3} + \cdots + Q(t_1 +t_2 -1) \nonumber \\
           =  \sum_{t_3 = t_1+1}^{t_1 + t_2 -1} Q(t_3) p^{t_1 + t_2 -t_3 -1}  
\end{equation} 
The first series on the RHS of equation (13) accounts for the situation where 
the LV hops out of the site $j=1$ before the FV arrives at the site $j=0$ 
and the second series accounts for the situation where the FV arrives at 
$j=0$ before the LV hops out of the site $j=1$.  
The general term in equation~(15) means that the LV halts for 
$t_3$ time-steps ($t_3 > t_1$), the FV being blocked at site 
$j=0$ for ($t_3 - t_1$) time-steps, contributing only a factor of $Q(t_3)$. 
After the LV hops out of site $j=1$ (at $t=t_3$), the 
FV continues to halt at site $j=0$ upto $t=t_1+t_2$ by 
braking, picking up a factor $p^{t_1 + t_2 -t_3 -1}$.

Using the expression (11) for $Q(t_3)$ in (14) and (15) we get 
\begin{equation}
S(t_1)  = 1 - \left[ {\cal C}(\underline{1}|1)q + p \right]^{t_1}
\end{equation}
and    
\begin{equation}
R(t_1,t_2) = \frac{{\cal C}(\underline{1}|0)}{{\cal C}(\underline{1}|1)}\left[ {\cal C}(\underline{1}|1)q + p \right]^{t_1}
                   \left[ p^{t_2 - 1} - \left[ {\cal C}(\underline{1}|1)q + p \right]^{t_2 - 1} \right]
\end{equation}

So, finally, using (16) and (17) in (13) the total probability distribution 
${\cal P}(t)$ can be written as
\begin{eqnarray}
         {\cal P}(t) 
         = \left[q \frac {{\cal C}(1|\underline{0})}
          {{\cal C}(0|\underline{0})} \right]B^{t-1} +
          \left[q \frac {{\cal C}(\underline{1}|0)}
          {{\cal C}(\underline{1}|1)} \right]A^{t-1} \nonumber \\
          + \left[ \frac {q^2 A}{p-AB} \frac {{\cal C}(1|\underline{0})}
          {{\cal C}(\underline{1}|1)} - q \frac {{\cal C}(\underline{1}|0)}
          {{\cal C}(\underline{1}|1)} \right] \left(AB \right)^{t-1} \nonumber \\
          - \left[\frac {q^2 A}{p-AB} \frac {{\cal C}(1|\underline{0})}
          {{\cal C}(\underline{1}|1)} + q \frac {{\cal C}(1|\underline{0})}
          {{\cal C}(0|\underline{0})}\right] p^{t-1}
       \end{eqnarray}
       where $A = 1 - q{\cal C}(\underline{1}|0), 
       \quad B= 1 - q{\cal C}(1|\underline{0})$ and
       equation~(18) was obtained using equations~(3)--(6) and y is 
       given by eq.~(7). 

The equation (18) is in excellent agreement with the corresponding results 
of our computer simulation (see figs.1 (a) and (b)). Moreover, the 
distribution for the densities $c$ and $1-c$ are identical; this is a 
conseqence of the particle-hole symmetry in the problem when $V_{max}$ is 
unity. 

One of the simplest dynamical models of interacting particle systems is 
the so-called {\it asymmetric simple exclusion process} (ASEP) [6] which is 
sometimes regarded as a caricature of vehicular traffic; in this model   
one particle is picked at random and moved forward by one lattice site 
if the new site is empty. In this random sequential update, it is the 
random picking that introduces stochasticity (noise) into the model. 
Although speed of the particles do not explicitly enter 
into the rules, the effective speed of a particle in the ASEP can take 
only two values, namely, $0$ and $V_{max} = 1$. Since mean-field theory 
is known to give exact results for ASEP, the time-headway distribution 
for ASEP can be obtained from eq.(18) by using the mean-field approximations 
${\cal C}(\underbar{1}|1) = {\cal C}(\underbar{0}|1) = c  
= {\cal C}(1|\underbar{1}) = {\cal C}(1|\underbar{0})$ and 
${\cal C}(\underbar{1}|0) = {\cal C}(\underbar{0}|0) = 1-c  
= {\cal C}(0|\underbar{1}) = {\cal C}(0|\underbar{0})$.

We present our numerical data for the time-headway distribution in the 
NS model with $V_{max} = 5$ in fig.2 as several earlier works have 
demonstrated that this particular choice of $V_{max}$ leads to quite realistic 
qualitative descriptions of some other features of vehicular traffic on 
highways.  The qualitative features of the distribution in fig.2 are similar 
to those in fig.1, except that ${\cal P}(t=0) = 0$ 
but ${\cal P}(t=1)$ need not vanish when $V_{max} = 5$ in contrast to the fact 
that ${\cal P}(t \leq 1) = 0$, irrespective of the density of the vehicles, 
when $V_{max} = 1$. 

We have plotted the most probable time-headway $T_{mp}$ as a function of 
the density in fig.3; while the curve is symmetric on the two sides of 
$c = 1/2$ when $V_{max} = 1$ no such symmetry is exhibited for $V_{max} > 1$. 
The common qualitative trend of variation of $T_{mp}$ is consistent with 
one's intuitive expectation that both at very low and very high densities 
there are long time gaps in between the departures of two successive vehicles 
from a given site. 

Finally, the time-headway distributions and the trend of their variation 
with density are in qualitative agreement with the coresponding empirical 
data [2]. The time headway distribution in the NS model can be approximated 
well with the forms ${\cal P}(t) \propto (t - \tau)^{\lambda} e^{-\mu t}$
with $\tau = 1$ for $V_{max} = 1$ and $\tau = 0$ for $V_{max} = 5$ where the 
parameters $\lambda$ and $\mu$ depend on the density of the vehicles [7]. 
Our results demonstrate that, in spite of being only a minimal model, the 
NS model captures the essential qualitative features of the time-headway 
distribution of vehicular traffic on highways.

{\bf Acknowledgements:} 

One of the authors (DC) thanks D. Stauffer for useful comments and the 
Humboldt Foundation, for partial support through a research equipment grant. 

\newpage

\noindent{\bf References} 

\noindent [1] D.E. Wolf, M. Schreckenberg and A. Bachem (eds.) {\sl Traffic and Granular Flow} (World Scientific, 1996) 

\noindent [2] A.D. May, {\sl Traffic Flow Fundamentals} (Prentice-Hall, 1990) 

\noindent [3] K. Nagel and M. Schreckenberg, {\it J. Physique I}, {\bf2},
2221 (1992).

\noindent [4] M. Schreckenberg, A. Schadschneider, K. Nagel and N. Ito,
{\it Phys.Rev.E}, {\bf 51}, 2939 (1995).

\noindent [5] D. Chowdhury, A. Majumdar, K. Ghosh, S. Sinha and R.B. Stinchcombe, Submitted for publication. 

\noindent [6] For a review, see B. Schmittmann and R.K.P. Zia, {\sl Statistical Mechanics of Driven Diffusive Systems}, vol.17 of Phase Transitions and Critical Phenomena, eds. C. Domb and J.L. Lebowitz (Academic Press, 1995).

\noindent [7] K. Ghosh, A. Majumdar and  D. Chowdhury, to be published.

\newpage
\noindent{\bf Figure Captions:} 

\noindent{\bf Fig.1:} The time-headway distribution in the steady-state of 
the NS model with $V_{max} = 1$ ($p = 0.5$) for the densities 
(a) $c = 0.1 (+)$, $c = 0.25 (\times)$ and $c = 0.5$ $(\ast)$. 
(b) $c = 0.9 (+)$, $c = 0.75 (\times)$ and $c = 0.5$ $(\ast)$. 
The continuous curves represent the analytical result while the discrete 
data points have been obtained from computer simulation. 

\noindent{\bf Fig.2:} The time-headway distribution in the steady-state of 
the NS model with $V_{max} = 5$ ($p = 0.5$) for the densities $c = 0.10 (+), 
0.25 (\times), 0.50 (\ast), 0.75 (\Box), 0.90 (\rule{2mm}{2mm})$.   
The discrete data points have been obtained from computer simulation while 
the continuous curves are merely guides to the eye. 

\noindent{\bf Fig.3:} The most probable time-headway plotted as a function 
of the density of the vehicles in the steady-sate of the NS model. The full 
line corresponds to $V_{max} = 1$ and has been obtained from the expression 
(18); the corresponding data obtained from our computer simulation have been 
represented by the symbol $+$. The discrete data points (represented by the 
symbol $\times$) correspond to $V_{max} = 5$ and have been obtained from 
computer simulation; the dotted line joining these data points serves merely 
as a guide to the eye.

\end{document}